\documentclass[useAMS,usenatbib]{mn2e}

\usepackage[dvips]{graphicx}
\usepackage[]{txfonts}
\def\simless{\mathbin{\lower 3pt\hbox
{$\rlap{\raise 5pt\hbox{$\char'074$}}\mathchar"7218$}}}   
\def\simmore{\mathbin{\lower 3pt\hbox
{$\rlap{\raise 5pt\hbox{$\char'076$}}\mathchar"7218$}}}   
\newcommand{\be}{\begin{equation}}
\newcommand{\ee}{\end{equation}}
\title[Radio transients from stellar tidal disruptions]{Radio
  transients from stellar tidal disruption by massive black holes}
\author[D. Giannios and B. D. Metzger]
{Dimitrios Giannios$^{1}$\thanks{E-mail: giannios@astro.princeton.edu, bmetzger@astro.princeton.edu},
  Brian D. Metzger$^{1,2}$\\
$^{1}$Department of Astrophysical Sciences, Peyton Hall, Princeton
  University, Princeton, NJ 08544, USA\\
$^{2}$NASA Einstein Fellow\\}

\begin{document}
\date{Received / Accepted}
\pagerange{\pageref{firstpage}--\pageref{lastpage}} \pubyear{2011}

\maketitle

\label{firstpage}

\begin{abstract}

The tidal disruption of a star by a supermassive black hole provides us with a rare glimpse of these otherwise dormant beasts.  It has long been predicted that the disruption will be accompanied by a thermal `flare', powered by the accretion of bound stellar debris.  Several candidate disruptions have been discovered in this manner at optical, UV and X-ray wavelengths.  Here we explore the observational consequences if a modest fraction of the accretion power is channeled into an ultra-relativistic outflow.
We show that a relativistic jet decelerates due to its interaction with the interstellar medium at sub-parsec distances from the black hole.  Synchrotron radiation from electrons accelerated by the reverse shock powers a bright radio-infrared transient that peaks on a timescale $\sim 1$ yr after disruption.  Emission from the forward shock may be detectable for several years after the peak. Deep radio follow-up observations of tidal disruption candidates at late times can test for the presence of relativistic ejecta.  Upcoming radio transient surveys may independently discover tens to hundreds of tidal disruptions per year, complimenting searches at other wavelengths.  Non-thermal emission from tidal disruption probes the physics of jet formation under relatively clean conditions, in which the flow parameters are independently constrained.    

\end{abstract} 
  
\begin{keywords}
black hole physics -- galaxies: nuclei.
\end{keywords}

\section{Introduction} 
\label{intro}

Supermassive black holes (SMBHs) are most easily studied when they accrete at high rates for extended periods of time and power active galactic nuclei (AGN).  However, the majority of galactic nuclei are relatively quiet.  With the exception of Sgr A$^{\star}$ and a handful of nearby low luminosity AGN, underfed SMBHs are difficult to study.  Even in bright AGN, obtaining a detailed understanding of the accretion process is hindered by our incomplete knowledge of the `boundary conditions' of the flow, such as the accretion rate and the magnetic field strength.

A rare glimpse into the properties of normally quiescent SMBHs is afforded when a star passes sufficiently close that it is torn apart by the SMBH's tidal gravitational field (Hills 1975; 
Rees 1988; Goodman $\&$ Lee 1989).  Analytic estimates and numerical calculations show that the process of disruption leaves a significant fraction of the shredded star gravitationally bound to the black hole (e.g.~Rees 1988; Ayal et al.~2000).  The accretion of this stellar debris has long been predicted to power a thermal `flare' at optical, UV, and X-ray wavelengths that lasts for months to years (e.g.~Ulmer 1999; Stubbe $\&$ Quataert 2009, 2010).
  
Models of the stellar dynamics in galactic nuclei estimate that the tidal disruption rate is $\sim 10^{-3}-10^{-5}$yr$^{-1}$ per galaxy (Magorrian \& Tremaine 1999; Wang $\&$ Merritt 1004).  Although tidal disruptions are rare, and potentially difficult to detect in the bright and dust-extincted central regions of galaxies, surveys at X-ray (e.g.~Donley et al.~2002; Esquej et al.~2008), optical (van Velzen et al.~2010) and far UV wavelengths (Gezari et al.~2006, 2008, 2009) have now detected $\sim 10$ candidates (see Gezari 2009 for a recent review).  Future wide-field surveys, particularly at optical wavelengths (e.g.~the Large Synoptic Survey Telescope), have the potential to detect hundreds of tidal disruptions per year (Strubbe $\&$ Quataert 2009).  A detailed census of these events would provide a unique probe of SMBH demographics and the shape of the gravitational potential in galactic nuclei (e.g.~Merrit $\&$ Poon 2004).

In this Letter we consider a different observational signature of tidal disruption that occurs if the accretion of stellar debris powers a transient relativistic jet (\S\ref{sec:disruption}).  We show that if a modest fraction of the accretion power is used to accelerate material to ultra-relativistic speeds, the energy that is released when the jet interacts with the interstellar medium ($\S\ref{sec:interaction}$) may power bright non-thermal synchrotron emission at radio-infrared wavelengths that peaks on a timescale $\sim$ 1 year after disruption ($\S\ref{sec:emission}$).  We explore the prospects for detecting such radio flares in $\S\ref{sec:detection}$ and discuss our results in $\S\ref{sec:discussion}$.  

\section{Stellar Disruption and Accretion}
\label{sec:disruption}

Disruption occurs when the tidal force applied by the SMBH overcomes the self gravity of the star.  This occurs when the radius of orbital pericenter $R_{\rm p}$ is less than a critical distance (Rees 1988)
\be
R_{\rm t}\simeq (M_{\rm BH}/M_{\star})^{1/3}R_\star,
\ee
where $M_{\star}$ and $R_{\star}$ are the stellar mass and radius, respectively.  For black holes masses $M_{\rm BH} \gtrsim 10^8 M_{\odot}$ and solar-type stars with radii $R_{\star} \sim 10^{11}$ cm, disruption occurs inside the event horizon, i.e.~$R_{\rm t}\simless R_{\rm s} \approx 2GM_{\rm BH}/c^2$, such that no electromagnetic signature is expected.  For less massive black holes, disruption occurs outside $R_{\rm s}$.  However, the prevalence of {\it very} low mass SMBHs ($M_{\rm BH} \lesssim 10^{6}M_{\sun}$) in galactic nuclei is unclear (e.g.~Ferrarese et al.~2006; Greene $\&$ Ho 2007) and, even if they are abundant, their electromagnetic emission will be dim if it is limited to a fraction of the Eddington luminosity $L_{\rm Edd} \propto M_{\rm BH}$.  The `sweet spot' for bright tidal disruption, therefore, occurs for $M_{\rm BH}\sim 10^{7-8} M_{\odot}$.

Disruption unbinds $\sim 1/2$ of the stellar debris from the system (e.g.~Rees 1988).  The rest of the mass is placed onto highly eccentric orbits, which return it to the vicinity of the black hole on a wide range of timescales (Lacy et al. 1982; Evans \& Kochanek 1989; Laguna et al.~1993).  The most tightly bound material returns on a timescale (e.g. Strubbe \& Quataert 2009)
\be
t_{\rm fallback}\simeq \frac{2\pi}{6^{3/2}}\left(\frac{R_{\rm p}}{R_{\star}}\right)^{3/2}\left(\frac{R_{\rm p}^{3}}{GM_{\rm BH}}\right)^{1/2} \simeq 5 \left(\frac{M_{\rm BH}}{10^{7}M_{\sun}}\right)^{5/2}\quad {\rm days},
\ee
where the last expression is evaluated for a solar-type star ($R_{\star} = R_{\odot}$) with a pericenter distance $R_{\rm p} = 6GM_{\rm BH}/c^2\simless R_{\rm t}$.  For clarity we focus on this case throughout the remainder of this Letter.

As material returns to the SMBH, it shocks on itself, circularizes, and accretes (e.g.~Kochanek 1994).  The accretion rate peaks at $t \sim t_{\rm fallback}$ and declines $\propto t^{-5/3}$ thereafter\footnote{Note, however, that corrections to this power-law decline may occur at both early (Lodato et al. 2009) and late (Cannizzo et al.~1990) times.} (Rees 1988), viz.~
\be
\dot{M}=\frac{M_\star}{3t_{\rm fallback}}\left(\frac{t_{\rm
    fallback}}{t}\right)^{5/3} \approx 24 \left(\frac{M_{\rm BH}}{10^{7}M_{\sun}}\right)^{-5/2}\left(\frac{t_{\rm fallback}}{t}\right)^{5/3}M_{\sun}{\rm \,yr^{-1}}
\label{eq:mdot}
\ee
For $M_{\rm BH} \lesssim 4\times 10^{7}M_{\sun}$ the accretion rate is initially super-Eddington, i.e.~$\dot{M}>\dot{M}_{\rm Edd} \equiv 10L_{\rm Edd}/c^2 \approx 0.2(M_{\rm BH}/10^{7}M_{\sun})M_{\odot}$ yr$^{-1}$.  Accretion decreases below the Eddington rate on a timescale
\be
t_{\rm Edd}\simeq 0.24\left(M_{\rm BH}/10^7 M_{\sun}\right)^{2/5}{\rm yr}.
\label{eq:tedd}
\ee 
A $\sim 10^{7}M_{\odot}$ black hole with an accretion rate $\dot{M} \sim \dot{M}_{\rm Edd}$ produces thermal emission that peaks in the Far-UV ($kT \sim 30$ eV), but which may also be detectable at optical,
near-UV, and X-ray wavelengths.  Thermal `flares' with a characteristic duration $\sim t_{\rm Edd} \sim$ months are thus considered a hallmark signature of tidal disruption;  their blue optical-UV colors are the primary means by which they are currently identified (Gezari et al. 2009).
\vspace{-0.5cm}
\subsection{Relativistic Jet Production}
\label{sec:jet}

Black hole accretion is often accompanied by powerful, collimated relativistic outflows.  A comparison of the disk luminosity and jet power of blazars, for instance, shows that a substantial fraction of the accretion power goes into relativistic jets (e.g. Ghisellini et al. 2009).  A similar correlation between jet and accretion power is observed in elliptical galaxies (Churazov et al.~2002; Allen et al.~2006).  

Though common, powerful jets are not present in all accreting systems at all times.  In stellar mass X-ray binaries, for instance, jets are ubiquitous at low accretion rates $\dot{M} \lesssim 0.03\dot{M}_{\rm Edd}$ (the low-hard state), but they are intermittent (or absent entirely) at higher accretion rates (the `thermal/high' and `very high' states; Fender et al.~2004).  The typical radio `loudness' of AGN also increases for smaller Eddington ratios $\lambda \equiv \dot{M}/\dot{M}_{\rm Edd}$, but nevertheless some fraction of AGN show evidence for jets at all values of $\lambda$.  Furthermore, the `radio loudness dichotomy' (e.g.~Ho 2002; Sikora et al.~2007) suggests that a `second parameter' other than $\dot{M}$, such as the black hole spin (Wilson \& Colbert 1995) or the magnetic flux threading the disk (Spruit $\&$ Uzdensky 2005), may also control the jet power.

The total energy released by the accretion of a solar mass star is $E_{\rm acc } \sim 10^{53}$ erg for an assumed accretion efficiency $\sim 0.1$.  The fraction of this energy $\epsilon_{\rm j}$ placed into relativistic ejecta is highly uncertain and could vary between disruption events, depending e.g.~on the magnitude of the SMBH spin and its direction with respect to the angular momentum of the disk.  We leave $\epsilon_{\rm j}$ as a free parameter in our calculations, but adopt a fiducial value $ \epsilon_{\rm j} = 0.01$ as motivated below.  The total energy of the jet is thus $E_{\rm j}= \epsilon_{\rm j}E_{\rm acc}\sim 10^{51}(\epsilon_{\rm j}/0.01)$ erg, similar to that of a supernova or gamma-ray burst (GRB).  We furthermore assume that the jet has an opening angle $\theta_{\rm j} \approx 0.1$ and a bulk Lorentz factor $\Gamma_{\rm j}=10$, values typical of AGN jets.  

We consider two scenarios for producing a relativistic jet following tidal disruption.  The first assumes that a relativistic outflow is present during the early super-Eddington accretion phase ($\dot{M} > \dot{M}_{\rm Edd}$; $t < t_{\rm Edd} \sim 10^{7}$ s; eq.~[\ref{eq:tedd}]), and that during this time the jet kinetic luminosity is constant $L_{\rm j}= E_{\rm j}/t_{\rm j} $, where $t_{\rm j} \approx t_{\rm Edd}$ is the jet duration.  The jet efficiency can be written as $\epsilon_{\rm j} \sim 0.1(L_{\rm j}/L_{\rm Edd})(M_{\rm BH}/10^{7}M_{\sun})^{7/5}$, such that $\epsilon_{\rm j} = 0.01$ for $L_{\rm j} = 0.1L_{\rm Edd}$, $M_{\rm BH} \sim 10^{7}M_{\odot}$.  On one hand, this scenario is conservative because numerical simulations of super-Eddington accretion show that the accretion rate reaching the black hole can far exceed $\dot{M}_{\rm Edd}$ (e.g.~Ohsuga et al.~2005), in which case $L_{\rm j}$ could in principle be $\gg L_{\rm Edd}$.  On the other hand, whether a relativistic jet accompanies super-Eddington accretion is observationally uncertain. 

In a second, more conservative scenario we assume that a relativistic jet forms only once the accretion rate $\dot{M} \propto t^{-5/3}$ (eq.~[\ref{eq:mdot}]) decreases to $\lesssim 0.03 \dot{M}_{\rm Edd}\equiv \dot{M}_{\rm t}$, as occurs at times $t \gtrsim 10t_{\rm Edd}$.  When the accretion rate becomes this low, the accretion disk is predicted to become radiatively inefficient, geometrically thick, and particularly susceptible to outflows (e.g.~Narayan $\&$ Yi 1995; Blandford $\&$ Begelman 1999), a fact supported by observations of X-ray binaries in the low/hard state and the radio loudness of AGN at low Eddington ratios.  The total energy of the jet power in this scenario is $E_{\rm j} \sim 0.1\dot{M}_{\rm t}c^{2}t_{\rm j}$ where the jet `duration' is $t_{\rm j} \sim 10t_{\rm Edd}$.  The jet efficiency can be written $\epsilon_{\rm j} \approx 0.03(L_{\rm j}/0.1\dot{M}c^{2})(M_{\rm BH}/10^{7}M_{\sun})^{7/5}$.
\vspace{-0.5cm}
\section{Jet-ISM interaction}
\label{sec:interaction}

\begin{figure}
\resizebox{\hsize}{!}{\includegraphics[]{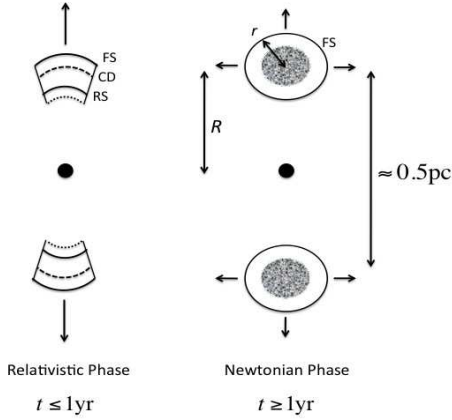}}
\caption[] {Sketch of the ultra-relativistic (left) and Newtonian (right) stages
  of jet deceleration. Initially the jet propagates radially, forming a forward/reverse shock (FS/RS) structure as it interacts with the ISM ($\S\ref{sec:relativistic}$).  Once the configuration has slowed to non-relativistic speeds, each of the two jet lobes expands quasi-spherically ($\S\ref{sec:nonrelativistic}$).  For a typical observer, emission peaks at the transition between the relativistic and non-relativistic stages due to synchrotron emitting electrons accelerated by the reverse shock (shown as grey shading on the right).} 
\label{fig:diagram}
\end{figure}

Jets from AGN are long-lived and propagate to $\sim $kpc-Mpc distances before dissipating their bulk energy (powering e.g.~giant radio lobes).  By contrast, any relativistic outflows from tidal disruption are transient and hence decelerate via interaction with the interstellar medium (ISM) at a much smaller distance $\lesssim$ pc.  As the jet drives a forward shock (FS) into the ISM, a reverse shock (RS) propagates through the ejecta, slowing it down ($\S\ref{sec:relativistic}$).  As discussed below, once the reverse shock crosses the jet, the entire configuration (swept-up ISM and shocked ejecta) has been slowed to mildly relativistic speeds.  After this point, the flow relaxes into a quasi-spherical, non-relativistic Sedov phase ($\S\ref{sec:nonrelativistic}$; see Fig.~\ref{fig:diagram}).

\vspace{-0.5cm}
\subsection{Relativistic Phase}
\label{sec:relativistic}

The interaction between a relativistic jet and the ISM has been studied analytically in two limits.  In the first approach, one considers a quasi-steady jet that slowly propagates through the ambient medium, creating a characteristic jet-cocoon structure (e.g.~Begelman $\&$ Cioffi 1989).  This limit is appropriate for long-lived AGN jets.  A second approach considers an impulsive ejection of relativistic material (a thin shell or `pancake') that produces a forward-reverse shock structure upon interacting with the ISM (Sari \& Piran 1995). This limit applies to GRB jets.   

Neither the steady nor the impulsive limits strictly apply to the jets from tidal disruption.  We nevertheless adopt the impulsive limit because the time required for the jet to appreciably decelerate $t_{\rm cr} \sim R_{\rm cr}/c$ (defined below) is $\simmore$ to the jet duration $t_{\rm j}$.  We furthermore approximate the intrinsically multidimensional problem as the one dimensional (1D radial) deceleration of a conical jet.  This approach is supported by relativistic hydrodynamical simulations (e.g.~Zhang \& MacFadyen 2009), which demonstrate that 2D effects, such as sideways expansion, become important only in the late, non-relativistic phases of the interaction ($\S\ref{sec:nonrelativistic}$).  

As the jet drives a FS into the ISM, a RS propagates back into the unshocked ejecta.  By assuming pressure balance at the contact discontinuity and solving the shock jump conditions, Sari \& Piran (1995) calculate the bulk Lorentz factor $\Gamma_{\rm sh}$ of the shocked fluid as a function of the radius $R$ of the flow.  For ISM of constant density $n_{\rm ISM}$ they show that while the reverse shock crosses the ejecta, the shocked fluid decelerates as $\Gamma_{\rm sh}(R) \simeq \Gamma_{\rm cr}(R/R_{\rm cr})^{-1/2}$, where
\be
R_{\rm cr}\simeq 2\Gamma_{\rm cr}^{2}ct_{\rm j} \approx 7 \times 10^{17}\left(\frac{\epsilon_{\rm j}}{0.01}\right)^{1/4}\left(\frac{n_{\rm ISM}}{10\,{\rm cm^{-3}}}\right)^{-1/4}\left(\frac{t_{\rm j}}{10^{7}\,{\rm s}}\right)^{1/4}\rm cm
\label{eq:Rcr}
\ee
is the radius at which the reverse shock fully crosses the ejecta, and
\be
\Gamma_{\rm cr}= \left(\frac{E_{\rm iso}}{64\pi n_{\rm ISM}m_p c^5 t_{\rm j}^{3}}\right)^{1/8} \simeq 1.1\left(\frac{\epsilon_{\rm j}}{0.01}\right)^{1/8}\left(\frac{n_{\rm ISM}}{10\,{\rm cm^{-3}}}\right)^{-1/8}\left(\frac{t_{\rm j}}{10^{7}\,{\rm s}}\right)^{-3/8}
\label{eq:gammacr}
\ee
is the Lorentz factor at $R \sim R_{\rm sh}$.  Here $E_{\rm iso} = E_{\rm j}f_{\rm b}^{-1}$ is the `isotropic' jet energy and $f_{\rm b} \equiv \theta_{\rm j}^{2}/2 \approx 5\times 10^{-3}$ is the jet beaming fraction.  We scale the ISM density to $\sim 10$ cm$^{-3}$, similar to that inferred from X-ray observations of the inner parsec of the Galactic centre (e.g.~Baganov et al.~2003; Quataert 2004).

Equations (\ref{eq:Rcr}) and (\ref{eq:gammacr}) show that the jet decelerates to a mildly relativistic velocity at $R_{\rm cr} \sim 0.1-1$ pc on a timescale $t_{\rm cr} \sim R_{\rm cr}/\beta_{\rm cr}c \sim 1$ and 3 years for $t_{\rm j} \sim t_{\rm Edd}$ and $t_{\rm j} \sim 10t_{\rm Edd}$, respectively.  Note that this result depends very weakly on the precise values of $\epsilon_{\rm j}$ and $n_{\rm ISM}$.  Importantly, {\it most of the bulk kinetic energy of the jet is dissipated by the reverse shock}, which decelerates the ejecta from its initial Lorentz factor $\Gamma_{\rm j} \sim 10$ to its final value $\Gamma_{\rm cr}\sim1$ with most of the dissipation occuring at radii $\sim R_{\rm cr}$.  Dissipation by the forward shock is sub-dominate because it is only mildly relativistic\footnote{In GRB jets the opposite holds true; the forward shock is relativistic and the reverse shock is mildly relativistic.  This is because the duration of the GRB $t_{\rm GRB}\ll t_{\rm j}$.} with $\Gamma_{\rm FS} \sim \Gamma_{\rm cr} \sim 1$ at $R \sim R_{\rm cr}$.  As we discuss in $\S\ref{sec:emission}$, emission from the jet is also dominated by the reverse shock.
 \vspace{-0.5cm}
\subsection{Non-Relativistic Phase}
\label{sec:nonrelativistic}

After the RS has crossed the jet ($R \gtrsim R_{\rm cr}$), the fluid expands mildly relativistically.  After a transient phase the outflow along each jet head relaxes into a non-relativistic, quasi-spherical expansion.  The evolution of each `lobe' at late times can thus be approximated by a Sedov-Taylor solution, for which the radial velocity evolves as $\beta_{\rm NR } \sim r^{-3/2}$, where a lower case $r$ denotes the radial distance measured from the center of each lobe (at radius $R \sim R_{\rm cr}$ from the black hole; see Fig.~\ref{fig:diagram} for an illustration).  During the non-relativistic phase the gas pressure decreases as $P \propto \beta^2_{\rm NR}\propto r^{-3}$, such that adiabatic expansion cools the electrons accelerated at the reverse shock (denoted by the shaded region in Fig.~\ref{fig:diagram}).  As discussed in $\S\ref{sec:lc}$, this controls the rate at which the RS emission declines after the peak.
  
\vspace{-0.5cm}
\section{Emission}
\label{sec:emission}

In calculating the synchrotron emission from the forward and reverse shocks, we make several standard assumptions regarding the shock microphysics (e.g. Sari et al. 1998), as motivated by the phenomenology of GRB afterglows.  First, we assume that a fraction $\epsilon_e\sim 0.1$ of the energy dissipated at the shock is used to accelerate relativistic electrons.  We assume the electron distribution is a power-law with index $p\sim 2.5$, such that the minimum electron Lorentz factor is $\gamma_{\rm m}=\epsilon_e (\Gamma_{\rm rel}-1)(p-2)m_p/(p-1)m_e \approx 600\epsilon_{e}(\Gamma_{\rm rel}-1)$, where $\Gamma_{\rm rel}$ is the relative Lorentz factor between the shocked and unshocked fluid.  We furthermore assume that the magnetic energy density of the shocked fluid is a fraction $\epsilon_B\sim 10^{-2}$ of the total energy density.\footnote{If the jet is strongly magnetized at $R \sim R_{\rm cr}$, then both the particle acceleration (e.g.~Sironi $\&$ Spitkovsky 2009) and dissipation at the reverse shock (Mimica et al.~2009) may be strongly suppressed.  However, since deceleration occurs at large radii $R_{\rm cr} \sim 10^{5.5} R_{\rm s}$, it is reasonable to assume that jet has fully accelerated, in which case the residual magnetization may be sufficiently low for a strong shock and non-thermal particle acceleration.}
\vspace{-0.5cm}
\subsection{Peak Emission}

In general the axis of the jet will be directed an angle $\theta_{\rm obs} \sim 1$ rad with respect to the line of site.  Emission is thus weak during the initial relativistic stage ($\S\ref{sec:relativistic}$) because (1) the luminosity is Doppler deboosted by a factor $\propto \delta^{4}$, where $\delta= \Gamma_{\rm sh}^{-1}(1-\beta_{\rm sh} \cos \theta_{\rm obs})^{-1}\sim 2/\Gamma_{\rm sh}\ll 1$ and (2) only a modest fraction of the jet has interacted at times $t \ll t_{\rm cr}$ ($R \ll R_{\rm cr}$).  Emission peaks when the reverse shock passes entirely through the ejecta ($t \sim t_{\rm peak} \sim t_{\rm cr}$; $R \sim R_{\rm cr}$) because a large fraction of the jet energy is dissipated around this time and the expansion has become mildly relativistic ($\delta \sim 1$). 

We now estimate the synchrotron emission at peak brightness.  Motivated by the above discussion, in what follows we assume that $\beta_{\rm sh} = 0.6$ $(\Gamma_{\rm sh} = 1.25)$ at $t \sim t_{\rm peak}$.  On timescales $t_{\rm peak} \sim t_{\rm cr}$, the relative Lorentz factor between the shocked and unshocked fluid is $\Gamma_{\rm rel,RS} \approx \Gamma_{\rm j}\Gamma_{\rm cr}(1-\beta_{\rm j}\beta_{\rm cr})\simeq \Gamma_{\rm j}/2$ and $\Gamma_{\rm rel,FS} = \Gamma_{\rm sh} $ for the RS and FS, respectively.  The minimum Lorentz factor of the accelerated electrons is $\gamma_{\rm m,FS} \simeq 20(\epsilon_{e}/0.1)$ and $\gamma_{\rm m,RS} \simeq 300(\epsilon_{e}/0.1)(\Gamma_{\rm j}/10)$.  Synchrotron flux peaks at the frequency
\begin{eqnarray}
\nu_{\rm m} \simeq \frac{eB\gamma_{\rm m}^{2}}{2\pi m_{e}c} \approx
 \left\{
\begin{array}{lr}
 0.1\left(\frac{\epsilon_{e}}{0.1}\right)^{2}\left(\frac{\epsilon_{B}}{0.01}\right)^{1/2}\left(\frac{n_{\rm ISM}}{10{\,\rm cm^{-3}}}\right)^{1/2}{\rm\,GHz}
&
{\rm (FS)} \\
 25\left(\frac{\epsilon_{e}}{0.1}\right)^{2}\left(\frac{\epsilon_{B}}{0.01}\right)^{1/2}\left(\frac{n_{\rm ISM}}{10{\,\rm cm^{-3}}}\right)^{1/2}\left(\frac{\Gamma_{\rm j}}{10}\right)^{2}{\rm\,GHz}
&
{\rm (RS)} \\
\end{array}
\right., 
\label{eq:num}
\end{eqnarray}  
where $B=\sqrt{8\pi e_{\rm th}\epsilon_{B}}$ and $e_{\rm th} \simeq (4\Gamma_{\rm sh}+3)(\Gamma_{\rm sh}-1)m_{p}n_{\rm ISM}c^2 \simeq 2m_{p}n_{\rm ISM}c^2$ are the magnetic field strength and energy density of the shocked fluid, respectively (e.g.~Sari $\&$ Piran 1995).

The synchrotron luminosity at $\nu_{\rm m}$ is $L_{\rm tot} \simeq (4/3)N_e\sigma_T c\gamma_{\rm m}^2 U_B$, where $N_e$ is the number of radiating electrons and $U_B = B^{2}/8\pi$.  At $t_{\rm peak} \sim t_{\rm cr}$ the number of ISM electrons swept up by the FS $N_{\rm e,FS}=2\pi\theta_{\rm j}^2 R_{\rm
  cr}^3n_{\rm ISM}/3\sim 10^{53}$ is comparable to the number in the shocked jet $N_{\rm e,RS}=E_{\rm j}/m_{p} c^2\Gamma_{\rm j} \sim
7\times 10^{52}(\epsilon_{\rm j}/0.01)(\Gamma_{\rm j}/10)^{-1}$.  The total emission from the RS thus exceeds that from the FS by a factor $\sim (\gamma_{\rm m,RS}/\gamma_{\rm m,FS})^{2} \gtrsim 10^{2}(\Gamma_{\rm j}/10)^2$.  Although naively the FS might still appear to dominate at low frequencies, self-absorption suppresses the emission at $\nu \lesssim 1$ GHz. The contribution of the FS is modest at the peak emission. The synchrotron luminosity $L_{\rm tot}$ of the RS is
\be 
L_{\rm tot} \approx 6\times 10^{40}(\epsilon_{\rm j}/0.01)(\epsilon_{e}/0.1)^{2}(\epsilon_{B}/0.01)(\Gamma_{\rm j}/10)\quad{\rm erg\,s^{-1}}.
\label{eq:Ptot}
\ee
The peak flux at $\nu \approx \nu_{\rm m}$ for an event at distance $D$ is thus given by
$F_{\nu_{\rm m}} \simeq {L_{\rm tot}}/{4\pi D^2 \nu_{\rm m}}$ or
\be 
F_{\nu_{\rm m}} \simeq 2\left(\frac{\epsilon_{\rm j}}{10^{-2}}\right)\left(\frac{\epsilon_{B}}{10^{-2}}\right)^{1/2}\left(\frac{n_{\rm ISM}}{10{\rm\,cm^{-3}}}\right)^{1/2}\left(\frac{\Gamma_{\rm j}}{10}\right)^{-1}\left(\frac{D}{\rm Gpc}\right)^{-2}{\rm mJy}, 
\label{eq:Fpeak}
\ee
Note that if the jet is modestly energetic ($\epsilon_{\rm j} \sim 10^{-3}-10^{-2}$), the predicted flux $\sim 0.1-1$ mJy is well within the sensitivity of current radio telescopes at $\sim $Gpc distances (redshift $z \approx 0.2$) similar to those of candidate disruption events. 

\begin{figure}
\resizebox{\hsize}{!}{\includegraphics[angle=270]{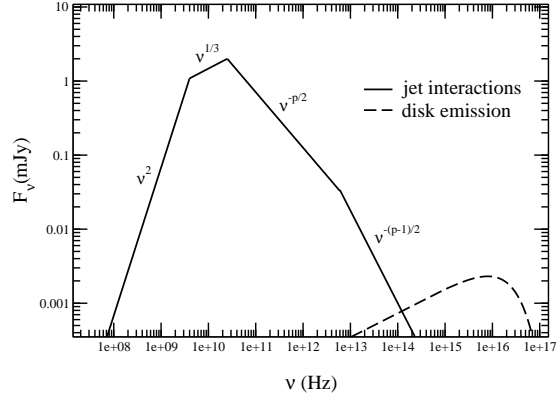}}
\caption[] {Predicted synchrotron spectrum ({\it solid line}) at the time of peak emission $t_{\rm peak} \sim t_{\rm cr} \sim 1$ year from the tidal disruption of a solar mass star by a $10^7M_{\odot}$ black hole at a distance $D=1$ Gpc.  From low to high frequencies, the three breaks correspond to the self absorption, characteristic and cooling frequency, respectively.  The dashed line shows an estimate of the thermal emission from the disk at a similar epoch.}
\label{fig:spectrum}
\end{figure}

Figure \ref{fig:spectrum} shows the spectrum of the RS emission at $t \approx t_{\rm peak} \sim t_{\rm cr}$.  In addition to the synchrotron peak frequency at $\nu_{\rm m} \approx $ 30 GHz (eq.~[\ref{eq:num}]), the two other breaks denote the self-absorption and cooling frequencies (at lower and higher frequencies, respectively).  Self absorption occurs below the frequency at which the optically-thin synchrotron luminosity equals the Rayleigh-Jeans luminosity of a plasma with temperature $T_e= \gamma_{\rm e,RS}m_{e}c^{2}/3$.  The cooling frequency is the characteristic synchrotron frequency of an electron with a cooling timescale equal to the expansion timescale $\sim t_{\rm cr}$. 

In Figure \ref{fig:spectrum} we plot for comparison an estimate of the contemporaneous emission from the accretion disk (assuming an extended thin disk).  At $t \sim t_{\rm peak} \sim 1$ year the disk luminosity is $L_{\rm d} \sim 0.1\dot{M}c^{2}|_{t_{\rm peak}} \sim 10^{44}$ erg s$^{-1}$ for $M_{\rm BH} \sim 10^{7}M_{\odot}$, i.e.~a factor $\sim 10^{3}$ brighter than the that of the RS for fiducial parameters (eq.~[\ref{eq:Ptot}]).  However, because the disk emission peaks in the Far-UV, emission from the jet dominates at radio-IR wavelengths.  Although emission from the disk appears to dominate that from the jet at optical-UV wavelengths, the opposite may be possible for more realistic disk models.  We note that the light curves of a few tidal disruption candidates `flatten' on a timescale $t \sim 1$ year (e.g.~D1-9, D3-13; Gezari et al.~2008, 2009).  If this flattening is due to the contamination of emission from the jet, we predict that the optical-UV spectrum should simultaneously flatten to $\nu F_{\nu} \sim$ const (see also Wong et al. 2007).
 \vspace{-0.5cm}
\subsection{Light Curve Rise and Decay}
\label{sec:lc}

In this section we estimate how the light curve rises prior to, and declines following, the epoch of peak emission ($t \sim t_{\rm peak} \sim t_{\rm cr}$).  Prior to the peak $t \ll t_{\rm cr}$, the RS is still crossing the jet, and the Lorentz factor of the shocked fluid decreases as $\Gamma_{\rm sh} \propto R^{-1/2}$.  The characteristic synchrotron frequency and the radiated power scale as $\nu_{\rm m} \propto \delta B \gamma_{\rm m}^{2}$ and $L_{\rm tot} \propto \delta^{4}N_{e}\gamma_{\rm e}^{2}B^{2}$, respectively.  For a typical observer angle $\theta_{\rm obs} \sim 1$ rad, the Doppler factor evolves as $\delta \propto \Gamma_{\rm sh}^{-1/2} \propto R^{1/2}$, while the time measured by the observer is $t_{\rm obs} \sim (R/\beta_{\rm sh}c)(1-\beta_{\rm sh}\cos\theta_{\rm obs})\sim R/2c\propto R$.  The energy density of the shocked fluid $e_{\rm th} \propto B^{2}$ scales as $\propto \Gamma_{\rm sh}^{2} \propto R^{-1}$, the minimum electron Lorentz factor $\gamma_{\rm m} \propto \Gamma_{\rm j}/\Gamma_{\rm sh} \propto R^{1/2}$, and the number of electrons swept up by the reverse shock evolves as $N_{e} \propto \int(\beta_{\rm j}-\beta_{\rm sh})dR \sim R_{\rm sh}/\Gamma_{\rm sh}^{2} \propto R_{\rm sh}^{2}$.  Combining these results, we find that $\nu_{\rm m} \propto R \propto t_{\rm obs}$, $L_{\rm tot} \propto R^{4} \propto t_{\rm obs}^{4}$, and $F_{\rm m} \propto L_{\rm tot}/\nu_{\rm m} \propto t_{\rm obs}^{3}$.  This steep predicted rise in the radio flux suggests that observations at times $\ll t_{\rm peak} \sim 1$ yr might be too early to detect the radio emission.

Once the RS crosses the jet, only the FS continues to accelerate electrons.  Nevertheless, for some period after the peak, the emission is still dominated by the electrons originally accelerated by the reverse shock when $R \sim R_{\rm cr}$, which adiabatically cool as the outflow expands.  Once the fluid transitions to a non-relativistic Sedov-Taylor expansion, its radius increases as $r \propto t^{2/5}$ and its pressure decreases as $P \propto r^{-3}$; the temperature of the relativistic electron `gas' $T_{e} \propto \gamma_{\rm m}$ thus decreases $\propto P^{1/4} \propto r^{-3/4}$.  The magnetic field generated by the shock decreases $B^{2} \propto P \propto r^{-3}$.  Combining these results, one finds that $\nu_{\rm m} \propto t_{\rm obs}^{-6/5}$, $L_{\rm tot} \propto t_{\rm obs}^{-9/5}$, and $F_{\rm m} \propto t_{\rm obs}^{-3/5}$.  This relatively steep decay, combined with the steep rise predicted above, suggests that the duration of the peak $\delta t$ is $\ll t_{\rm peak}$.  Hydrodynamic simulations of the trans-relativistic stage are, however, required to more precisely quantify the light curve shape.

Although RS electrons dominate the emission immediately after $t_{\rm peak}$, synchrotron radiation from the `fresh' electrons accelerated by the FS will take over at later times.  A useful analogy in this case are the radio afterglows of nearby GRBs.  For GRB 030329, the redshift $z \simeq 0.17$ and total energy in relativistic ejecta $\sim 3\times 10^{50}$ ergs are similar to the characteristic values for tidal disruption.  The 2.3 GHz flux of 030329 was $F_{\nu} \sim$ mJy at $t\sim 1$ year after the burst, with the light curve decaying relatively gradually ($F_{\nu} \propto t^{-1}$; van der Horst et al.~2008).  
\vspace{-0.5cm}
\section{Detection Prospects}
\label{sec:detection}

One method to detect transient radio emission from tidal disruption is via follow-up observations of candidate events detected at other wavelengths (e.g.~optical, UV and X-rays).  The only relevant radio follow-up to date (of which we are aware) is of the SDSS Stripe 82 candidate TDE2 discovered by van Velzen et al.~(2010), for which VLA observations at $\nu = 8.5$ GHz place upper limits of $F_{\nu} \lesssim 0.2$ mJy at $t \sim 7$ and $\sim 90$ days after discovery.  Unfortunately, these non-detections are not constraining because the flux is predicted to rise rapidly $\propto t^{3}$ prior to the peak at $t\sim 1$ year ($\S\ref{sec:lc}$).  

Radio emission might still be detectable {\it now} from candidates discovered in the past few years (e.g.~Gezari et al.~2007, 2008, 2009). Moreover, for a typical source redshift $z \sim 0.2$, the GHz emission from a tidal disruption could be sufficiently bright ($\sim$ mJy; eq.~[\ref{eq:Fpeak}]) and extended that the diametric radio lobes (Fig.~\ref{fig:diagram}) physical (angular) separation $\sim 2R_{\rm cr} \sim 0.5$ pc ($\approx 0.1$mas), could be resolved using Very Long Baseline Interferometry, again analogous to the afterglow of GRB 030329 (Taylor et al.~2004).  

Another exciting prospect is the {\it discovery} of tidal disruptions with present or future radio surveys, such as Pi GHz Sky Survey ({\it PiGS}) of the Allen Telescope Array (Bower et al. 2010).  Given {\it PiGS} predicted field of view ($\sim 10,000$ deg$^{2}$), cadence ($\sim$ days-years) and sensitivity limit $\sim 5$ mJy at $\nu \sim 3.1$ GHz, we estimate that $\sim 100$ tidal disruptions could be detected per year, assuming fiducial values for the jet properties from equation (\ref{eq:Fpeak}); a tidal disruption rate $\sim 10^{-5}$ yr$^{-1}$ per galaxy; and a local black hole density $\sim 3\times 10^{-3}$ Mpc$^{-3}$ (e.g.~Hopkins et al.~2007).  Another possible strategy is a targeted survey towards nominally radio quiet spheroids or bulges, which are preselected to harbor black holes with masses $M_{\rm BH} \sim 10^{7}-10^{8} M_{\odot}$ (D.~Frail, private communication).

\vspace{-0.5cm}

\section{Discussion}
\label{sec:discussion}
In analogy to other accreting black hole systems such as AGN and X-ray binaries, the tidal disruption of a star by a SMBH may be accompanied by a powerful, relativistic jet.  Such jet may be responsible to the acceleration of ultra-high energy cosmic rays (Farrar \& Gruzinov 2009). In this Letter we have shown that the interaction of such a jet with the ambient ISM produces a bright radio-infrared transient that peaks on a timescale $\sim 1$ year after disruption.  Radio transients from tidal disruption may be detected blindly with upcoming radio transient surveys, or via follow-up observations of disruption events detected at other wavelengths on timescales of several months to years.  Upcoming radio transient surveys similar to {\it PiGS} could discover tens or hundreds of tidal disruptions per year.  The good radio localization would confirm the location of the transient at the center of the galaxy.  

The detection and characterization of tidal disruptions is a promising means to study the demographics of SMBHs (Gezari et al.~2009).  Tidal disruption also provides a unique venue to study the physics of accretion, and its connection to jet formation, under conditions in which the `boundary conditions' of the flow are relatively well-determined.  The thermal disk emission provides a direct estimate of the accretion rate.  Furthermore, because the magnetic field strengths of solar type stars are known, the magnetic flux through the accreting material is well constrained.  

An important unsolved question associated with jet formation is whether the magnetic field necessary to power the jet is advected with the flow (e.g. Spruit \& Uzdensky 2005), or whether it is generated locally in the disk by instabilities or dynamo action.  Because the magnetic flux of a solar-type star is insufficient to drive a powerful jet, the detection of bright radio emission associated with tidal disruption would favor the hypothesis that locally-generated fields are responsible for jet (and vice versa).  Furthermore, if only a fraction of (otherwise similar) tidal disruptions shows evidence for a jet, it would indicate that a second parameter in addition to the accretion rate (such as the black hole spin) controls the jet strength.  
\vspace{-1cm}
\section*{Acknowledgments}
We thank S.~Gezari, D.~Frail and L. Strubbe for helpful comments.  DG acknowledges support from the Lyman Spitzer, Jr. Fellowship awarded by the Department of Astrophysical Sciences at Princeton University.  BDM is supported by NASA through Einstein Postdoctoral Fellowship grant number PF9-00065 awarded by the Chandra X-ray Center, which is operated by the Smithsonian Astrophysical Observatory for NASA under contract NAS8-03060.

\end{document}